\documentclass{ws-mpla}
\usepackage[super]{cite}
\usepackage{graphicx}
\begin{document}

\markboth{M.A.Zubkov}
{Dynamical torsion as the microscopic origin of the neutrino seesaw}

\catchline{}{}{}{}{}

\title{Dynamical torsion as the microscopic origin of the neutrino seesaw.}

\author{\footnotesize M.A.Zubkov\footnote{
on leave of absence from ITEP, B.Cheremushkinskaya 25, Moscow, 117259, Russia.}}

\address{The University of Western Ontario, Department of Applied
Mathematics,
 1151 Richmond St. N., London (ON), Canada N6A 5B7 \\
zubkov@itep.ru}

\maketitle

\pub{Received (Day Month Year)}{Revised (Day Month Year)}

\begin{abstract}
It is assumed, that there are two scales in quantum gravity. Metric fluctuates at the scales of the order of the Plank mass. The second scale $M_T$ is related to the fluctuations of torsion. We suppose, that it may be as low as $M_T \sim 1$ TeV.
Due to the non - minimal coupling to torsion the attractive interaction between the fermions appears. The non - minimal coupling admits the appearance of different coupling constants for different fermions. This opens the possibility that the interaction with torsion gives the Majorana masses for the right - handed neutrinos (that are assumed to be of the order of $M_T$). We suppose, that the Dirac masses for the neutrino are all close to the mass of electron. This gives the light neutrino masses $\le  0.25$ eV. In addition, the model predicts the appearance of Majorons that may contribute to the dark matter as well as to the invisible decay of the $125$ GeV Higgs boson.
\end{abstract}

\ccode{12.60.Fr 04.50.Kd}

\newcommand{\revision}[1]{{#1}}
\newcommand{\revisionA}[1]{{#1}}
\newcommand{\revisionB}[1]{{#1}}

\bibliographystyle{apsrev}


\newcommand{\br}{{\bf r}}
\newcommand{\bu}{{\bf \delta}}
\newcommand{\bk}{{\bf k}}
\newcommand{\bq}{{\bf q}}
\def\({\left(}
\def\){\right)}
\def\[{\left[}
\def\]{\right]}

\newcommand{\barray}{\begin{eqnarray}}
\newcommand{\earray}{\end{eqnarray}}
\newcommand{\nn}{\nonumber \\}
\newcommand{\nl}{& \nonumber \\ &}
\newcommand{\bnl}{\right .  \nonumber \\  \left .}
\newcommand{\dbnl}{\right .\right . & \nonumber \\ & \left .\left .}

\newcommand{\beq}{\begin{equation}}
\newcommand{\eeq}{\end{equation}}
\newcommand{\ba}{\begin{array}}
\newcommand{\ea}{\end{array}}
\newcommand{\bea}{\begin{eqnarray}}
\newcommand{\eea}{\end{eqnarray} }
\newcommand{\be}{\begin{eqnarray}}
\newcommand{\ee}{\end{eqnarray} }
\newcommand{\bal}{\begin{align}}
\newcommand{\eal}{\end{align}}
\newcommand{\ei}{\end{itemize}}
\newcommand{\ben}{\begin{enumerate}}
\newcommand{\een}{\end{enumerate}}
\newcommand{\bc}{\begin{center}}
\newcommand{\ec}{\end{center}}
\newcommand{\bt}{\begin{table}}
\newcommand{\et}{\end{table}}
\newcommand{\btb}{\begin{tabular}}
\newcommand{\etb}{\end{tabular}}
\newcommand{\bvec}{\left ( \ba{c}}
\newcommand{\evec}{\ea \right )}

\newcommand\e{{e}}
\newcommand\eurA{\eur{A}}
\newcommand\scrA{\mathscr{A}}

\newcommand\eurB{\eur{B}}
\newcommand\scrB{\mathscr{B}}

\newcommand\eurV{\eur{V}}
\newcommand\scrV{\mathscr{V}}
\newcommand\scrW{\mathscr{W}}

\newcommand\eurD{\eur{D}}
\newcommand\eurJ{\eur{J}}
\newcommand\eurL{\eur{L}}
\newcommand\eurW{\eur{W}}

\newcommand\eubD{\eub{D}}
\newcommand\eubJ{\eub{J}}
\newcommand\eubL{\eub{L}}
\newcommand\eubW{\eub{W}}

\newcommand\bmupalpha{\bm\upalpha}
\newcommand\bmupbeta{\bm\upbeta}
\newcommand\bmuppsi{\bm\uppsi}
\newcommand\bmupphi{\bm\upphi}
\newcommand\bmuprho{\bm\uprho}
\newcommand\bmupxi{\bm\upxi}

\newcommand\calJ{\mathcal{J}}
\newcommand\calL{\mathcal{L}}

\newcommand{\notyet}[1]{{}}

\newcommand{\sgn}{\mathop{\rm sgn}}
\newcommand{\tr}{\mathop{\rm Tr}}
\newcommand{\rk}{\mathop{\rm rk}}
\newcommand{\rank}{\mathop{\rm rank}}
\newcommand{\corank}{\mathop{\rm corank}}
\newcommand{\range}{\mathop{\rm Range\,}}
\newcommand{\supp}{\mathop{\rm supp}}
\newcommand{\p}{\partial}
\renewcommand{\P}{\grave{\partial}}
\newcommand{\yDelta}{\grave{\Delta}}
\newcommand{\yD}{\grave{D}}
\newcommand{\yeurD}{\grave{\eur{D}}}
\newcommand{\yeubD}{\grave{\eub{D}}}
\newcommand{\at}[1]{\vert\sb{\sb{#1}}}
\newcommand{\At}[1]{\biggr\vert\sb{\sb{#1}}}
\newcommand{\vect}[1]{{\bold #1}}
\def\R{\mathbb{R}}
\newcommand{\C}{\mathbb{C}}
\def\hvar{{\hbar}}
\newcommand{\N}{\mathbb{N}}\newcommand{\Z}{\mathbb{Z}}
\newcommand{\Abs}[1]{\left\vert#1\right\vert}
\newcommand{\abs}[1]{\vert #1 \vert}
\newcommand{\Norm}[1]{\left\Vert #1 \right\Vert}
\newcommand{\norm}[1]{\Vert #1 \Vert}
\newcommand{\Const}{{C{\hskip -1.5pt}onst}\,}
\newcommand{\sothat}{{\rm ;}\ }
\newcommand{\Range}{\mathop{\rm Range}}
\newcommand{\ftc}[1]{$\blacktriangleright\!\!\blacktriangleright$\footnote{AC: #1}}



\newcommand{\const}{\mathop{\rm const}}

\renewcommand{\theequation}{\thesection.\arabic{equation}}

\makeatletter\@addtoreset{equation}{section}
\makeatother

\def\Tau{\mathcal{T}}

\def\os{{o}}
\def\ol{{O}}
\def\dist{\mathop{\rm dist}\nolimits}
\def\spec{\sigma}
\def\mod{\mathop{\rm mod}\nolimits}
\renewcommand{\Re}{\mathop{\rm{R\hskip -1pt e}}\nolimits}
\renewcommand{\Im}{\mathop{\rm{I\hskip -1pt m}}\nolimits}

\section{Introduction}
\label{1}

The recent discovery of the $125$ GeV scalar h - boson \cite{CMSHiggs,ATLASHiggs}, which shares many properties with the Higgs boson of the Standard Model (SM), confirms the existence of the Higgs mechanism for the formation of masses of $W$ and $Z$ bosons \cite{Englert,Higgs}. However, in spite of all the significance of this finding, the well - known questions related with the flavor physics remain open (in particular, the origin of the complicated spectra of quarks and leptons\footnote{
Besides, the Hierarchy problem points out the possibility, that the scalar Higgs bosons (and, in particular, the $125$ GeV h - boson itself) may be composite just like the Cooper pairs in superconductors are composite \cite{NJL}.}).  

In the present paper we consider the scenario in which the seesaw mechanism for the Majorana neutrino masses  originates in the interactions of the fermions with quantum gravity. Namely, we consider quantum gravity with torsion coupled non - minimally to fermions. \revisionB{Since the metric fluctuates at the scale of the order of the Plank mass, we ignore these fluctuations.  The second scale $M_T$ is related to the fluctuations of torsion. According to the present experimental limits, $M_T$ may be of the order of $1$ TeV or somewhat larger \cite{Shapiro,Shapiro1}. At such energies the metric field is frozen, and we are left with the gauge theory of the Lorentz group  \cite{Minkowski,Minkowski_}. The exchange by the gauge bosons of this group is able to provide an attractive interaction in two different channels. This may result both in the pairing between fermions and anti-fermions of different chiralities (a Dirac mass) and in the pairing of two fermions of the same chirality but with different directions of spin (a Majorana mass).} In principle, certain contributions to Dirac masses of fermions may appear in this way. However, we do not discuss here this possibility. Instead we concentrate on the possibility to generate the large Majorana masses for the right - handed neutrinos\footnote{Recently, the formation of the Dirac masses of the Standard Model fermions due to the interaction with torsion has been discussed in Refs. \cite{Z2013JHEP,Z2013JHEP1,chili}.
The idea that the interactions with the torsion are responsible for the dynamical electroweak symmetry breaking was first suggested in Ref. \cite{Z2010}. Recently this idea has been applied to the top - quark condensation pattern \cite{Miransky,Miransky1,Nambu,Marciano:1989xd,topcolor1,Hill:1991at,Hill:2002ap}
in Refs. \cite{Z2013JHEP,chili}. In another context, the formation of the four - fermion interactions due to the torsion that may lead to the condensation of fermions was considered in Refs. \cite{Freidel:2005sn,Randono:2005up,Mercuri:2006um,Alexandrov,Xue,Alexander1,Alexander2,shapiro2012}.
Mechanism that provides different values of the coupling constants for different flavors in the effective four - fermion interactions was not specified in Ref.\cite{Z2013JHEP}. In Ref. \cite{chili} it was suggested that the different values of the couplings are originated from the dimensional reduction of the $5D$ theory. In both papers, the Majorana masses for neutrinos were not considred. The crucial new point of the present paper is that Majorana masses may be generated.}.

According to the pattern considered in the given paper the major contributions to the $W$ and $Z$ - boson masses originate from the interaction with the $125$ GeV $h$ - boson. We do not discuss the mechanism that is responsible for the Dirac masses of SM fermions. It is related to the interaction with the $125$ GeV Higgs boson. Besides, there may be the contributions from the new inter - fermion interactions, and, in particular, from the one induced by torsion (as it was mentioned aboove).

The key observation is that the coupling constants entering the nonminimal interaction of the fermion fields with gravity can be different for different fermions (in particular, for the fermions with different chirality). The latter may result in the situation, when the interaction between the right - handed neutrinos is attractive while the interaction between the left - handed neutrinos is repulsive.
In the scenario considered in this paper, we assume that the Majorana masses $M_{\nu_R}$ of right-handed neutrinos
are close to each other and are of the order of $M_T$. We also imply, that all neutrinos acquire the Dirac masses of the order of the electron mass $m_{\nu} \sim m_e \approx 0.5$ MeV.  Then due to the seesaw mechanism \cite{seesaw,seesaw1,seesaw2,seesaw3,seesaw4}, we arrive at the light Majorana masses  $M^{light}_{\nu} \le 0.25$ eV. This allows to avoid the present experimental constraints on the neutrino masses both from the cosmological data and from the direct experiments \cite{neutrino}. \revisionB{Note that since in this scenario only the Standard Model (SM) fermions plus right-handed neutrinos are included, there are no additional difficulties related to the chiral anomalies (compare with the discussion in Ref. \cite{Shapiro}).}


It is worth mentioning, that previously there were attempts to explain the appearance of large Majorana masses for the right - handed neutrinos by the interaction with quantum gravity. In Ref. \cite{Barenboim:2010db} the exchange by ordinary graviton was discussed. It leads to the appearance of an extra large Majorana mass for the neutrino, which is much larger than the scale considered here. In Ref. \cite{Mavromatos:2012cc} the minimal interaction of fermions with torsion was discussed, the nonminimal coupling was not considered. As a result, the ordinary exchange by the quanta of torsion leads to the repulsive interaction in the Majorana channel, and the hypothetical appearance of the dynamical Majorana mass was related to some topological configurations.

The paper is organized as follows. In section \ref{Sectfermions} we discuss the details of the non - minimal coupling of fermion fields to quantum gravity. In section \ref{Sectmasses} we explain how the attractive interaction between the fermions is formed and how the fermion masses are induced.  \revisionB{In section \ref{concl} we summarize our results.}

\section{Weyl fermions in Riemann-Cartan space}
\label{Sectfermions}

\subsection{Definition}

{\it Throughout the text space - time indices are denoted by small Greek letters $\mu, \nu, \rho, ...$ while internal $SO(3,1)$ - indices are denoted by small Latin letters $a,b,c,...$. The Weyl spinor indices are denoted by large Latin letters $A,B,C,...$, the flavor indices are denoted by bold small Latin letters $\bf a,b,c,...$}
{\it Greek indices are lowered and lifted via metric tensor $g_{\mu \nu}$ while Latin indices are lowered and lifted by the metric tensor of Minkowski space $\eta_{ab}$. Internal $SO(3,1)$ indices are transformed to external space - time indices with the help of the vierbein $E_a^{\mu}$ and the inverse vierbein $E_{\mu}^a$.}

{The most general form of the action for the right - handed two - component Weyl spinor $\Psi_R$ in Riemann - Cartan space follows from the expression for massless Dirac spinor suggested in Ref. \cite{Alexandrov} (see also Refs. \cite{Diakonov,Diakonov1,Diakonov2}) and has the form:
\begin{eqnarray}
S_R & = & \frac{1}{2}\int E \Bigl( i \bar{\Psi}_R \sigma^{\mu} (1 + i B_R)
D_{\mu} \Psi_R + (h.c.) \Bigr) d^4 x, \label{Sf}
\end{eqnarray}
where $B_R$ is a real coupling constant, $E = {\rm det} E^a_{\mu}$, $\sigma^{\mu} =
E^{\mu}_a \sigma^a$, and while $\sigma^{a}$ for $a=1,2,3$ are the Pauli matrices, $\sigma^0$ is the unity matrix. The covariant derivative is denoted by $D_{\mu} =
\partial_{\mu} + \frac{1}{4}(\omega_{\mu}^{ab}+C_{\mu}^{ab})\sigma_{ab}$. Here for $a,b = 1,2,3$ we denote
$\sigma_{ab} = - \frac{1}{2}[\sigma_{a},\sigma_{b}]$ while $\sigma_{0b} = - \sigma_{b0} = \sigma_b$. The symbol  $(h.c.)$ means the hermitian conjugation (anticommuting variables $\Psi$ , $\bar{\Psi}$ are formally considered as mutually  conjugated operators.)}

 The torsion free spin
connection is denoted by $\omega_{\mu} $, while $C_{\mu}$ is the contorsion
tensor. They are related to $E^a_{\mu}$, the affine connection $\Gamma^{i}_{jk}$,
and torsion $T^a_{.\mu \nu}= T^{\rho}_{.\mu\nu}E^a_{\rho}$ as follows:
\begin{eqnarray}
0 &\equiv& \partial_{\nu}E^a_{\mu} - \Gamma^{\rho}_{\mu
\nu}E^a_{\rho} + \omega^{a}_{ . b\nu}E^b_{\mu}+ C^{a}_{ . b\nu}E^b_{\mu}\nonumber\\
\tilde{D}_{[\nu} E_{\mu]}^{a} &\equiv& \partial_{[\nu}E^a_{\mu]} +
\omega^{a}_{.b[\nu }E^b_{\mu]}=0\nonumber\\
T^a_{.\mu \nu} & \equiv & {D}_{[\nu} E_{\mu]}^{a} = \partial_{[\nu}E^a_{\mu]} +
\omega^{a}_{.b[\nu }E^b_{\mu]}+ C^{a}_{ . b[\nu}E^b_{\mu]}=C^{a}_{ .
b[\nu}E^b_{\mu]}
\end{eqnarray}

This results in:
\begin{eqnarray}
\{^{\alpha}_{\beta \gamma}\} & = &
\frac{1}{2}g^{\alpha\lambda}(\partial_{\beta}g_{\lambda
\gamma}+\partial_{\gamma}g_{\lambda \beta}-\partial_{\lambda}g_{\beta \gamma})\nonumber\\
C^{\rho}_{.\mu \nu} & = & \frac{1}{2}(T^{\rho}_{.\mu \nu}-T^{.\rho}_{\nu
.\mu}+T^{..\rho}_{\mu \nu})\nonumber\\
\Gamma^{\rho}_{\mu \nu}&=& \{^{\rho}_{\mu \nu}\} + C^{\rho}_{.\mu \nu}
\nonumber\\
\omega_{ab\mu} & = &\frac{1}{2}( c_{abc}-c_{cab}+c_{bca})E^{c}_{\mu}
\end{eqnarray}
Here $c_{abc} = \eta_{ad}E^{\mu}_b E_c^{\nu}\partial_{[\nu}E^d_{\mu]}$;
$T_{abc} = \eta_{ad}E^{\mu}_b E_c^{\nu}T_{.\mu\nu}^d$; $g_{\mu \nu} =
E^a_{\mu}E^b_{\nu}\eta_{ab}$; $\Gamma^{\rho}_{\mu \nu}-\Gamma^{\rho}_{\nu \mu}
= T^{\rho}_{.\mu \nu}$.

{The action for the left - handed spinor is obtained from Eq. (\ref{Sf}) by the change: \revisionB{$B_R \rightarrow B_L$, $\Psi_R \rightarrow \Psi_L$, $\sigma^a \rightarrow \bar{\sigma}^a$, where $\bar{\sigma}^0 = \sigma^0$ and $\bar{\sigma}^i = -\sigma^i$ for $i = 1,2,3$.}
The coupling constant $B_L$ may be different from $B_R$ if parity breaking is allowed.
The further step is to introduce different coupling constants for  different flavors. As a result we arrive at the two  Hermitian  $N_f\times N_f$ matrices $B_L$ and $B_R$, where $N_f$ is the number of the fermion flavors. Via  unitary transformations we can always make both $B_L$ and $B_R$ diagonal.}

\subsection{Expression through axial and vector torsion}

{Now let us introduce the irreducible tetrad components of torsion:
\begin{eqnarray}
S^{a}& =& \epsilon^{bcda}T_{bcd}\nonumber\\
T_{a}& =& T^{b}_{.ab}\nonumber\\
T_{abc}& =& \frac{1}{3}(T_{b} \eta_{ac}-T_{c}\eta_{ab}) -
\frac{1}{6}\epsilon_{abcd}S^{d} + q_{abc} \label{STT}
\end{eqnarray}
}
In terms of $S$ and $T$ the action for the fermions can be rewritten as:

\begin{eqnarray}
S_f & = & S_{\nabla} + \frac{1}{2}\int E\Bigl(
\frac{1}{4}S^d \Bigl(-\bar{\Psi}_L \bar{\sigma}_d \Psi_L + \bar{\Psi}_R \sigma_{d} \Psi_R\Bigr)\nonumber\\&&- T^d \Bigl(\bar{\Psi}_L \bar{\sigma}_d B_L \Psi_L + \bar{\Psi}_R B_R  \sigma_{d} \Psi_R\Bigr) \Bigr)d^4 x \label{Sf2222}
\end{eqnarray}
Here
\begin{equation}
S_{\nabla} = \frac{1}{2}\int E \Bigl(i\bar{\Psi}_L \bar{\sigma}^{\mu}
(1 + i B_L)
\tilde{D}_{\mu}\Psi_L  + i\bar{\Psi}_R \sigma^{\mu}
(1 + iB_R)
\tilde{D}_{\mu} \Psi_R  + (h.c.) \Bigr)d^4 x,\label{Ssimple0}
\end{equation}
\revisionB{where} $\tilde{D}_{\mu}$ is the torsion - free covariant derivative of general relativity.

Notice that in the particular case of flat metric $E^a_{\mu} = \delta^a_{\mu}$ the kinetic term is reduced to
\begin{equation}
S_{\nabla} = \frac{1}{2}\int \Bigl(i\bar{\Psi}_L \bar{\sigma}^{\mu}
\nabla_{\mu} \Psi_L  + i\bar{\Psi}_R \sigma^{\mu}
\nabla_{\mu} \Psi_R  + (h.c.) \Bigr)d^4 x,\label{Ssimple}
\end{equation}
where $\nabla_{\mu}$ is the usual derivative. For the transition from Eq. (\ref{Ssimple0}) to Eq. (\ref{Ssimple}) we take into account that in the flat case $\tilde{D}_{\mu} = \nabla_{\mu}$ does not contain spin connection and, therefore, commutes with $\sigma^{a},\bar{\sigma}^a$. As a result the hermitian conjugate expression denoted by $(h.c.)$ differs from the first term by the change $(1+iB_{L,R}) \rightarrow (1-i B_{L,R})$. In the resulting sum the coefficients $B_L$ and $B_R$ disappear.

\section{Appearance of fermion masses}
\label{Sectmasses}

\subsection{Effective four - fermion interaction due to torsion}

{The most general form of the action up to the terms quadratic in torsion, its derivatives, and in curvature was considered in \cite{Diakonov2}. In principle, several different torsion - dependent terms may appear even without derivatives of torsion.} { The most general action quadratic in torsion has the form \cite{Diakonov2}:
\begin{eqnarray}
S_T =    M_T^2 \int E\{ \frac{2 \chi_{VV}}{3}T^2 - \frac{\chi_{AA}}{24}S^2 + \chi_{AV}  ST \}d^4x  + \tilde{S}[q,E],\label{S0}
\end{eqnarray}
where $M_T$ is the dimensional parameter that gives the torsion fluctuation scale, $\chi_{VV}, \chi_{AA}, \chi_{AV}$ are dimensionless parameters, while $\tilde{S}$ depends on $q$ and $E$ but does not depend on $S, T$.
Below for the simplicity we consider the particular case, when $\chi_{AA} = \chi_{VV} = 1, \chi_{AV}=0$. The generalization of our consideration for the case of arbitrary $\chi_{AA}, \chi_{VV}, \chi_{AV}$ is straightforward. Our low energy action for torsion appears from the "minimal" prescription (the terminology of \cite{Diakonov2}). It corresponds to the choice of the so - called Palatini action
 \cite{Mercuri:2006um}:
\begin{eqnarray}
S_T =  -M_T^2 \int E E_a^{\mu}E_b^{\nu}G^{ab}_{\mu\nu} d^4x = M_T^2 \int E\{- R + \frac{2}{3}T^2 - \frac{1}{24}S^2  \}d^4x  + \tilde{S}[q,E]\label{S}
\end{eqnarray}
}{Here  $G^{ab}_{..\mu\nu} = [D_{\mu},D_{\nu}]^{ab}$ is the $SO(3,1)$ curvature,  $R$ is Riemannian scalar curvature. It is worth mentioning, that the negative sign at the term with $S^2$ in Eq. (\ref{S}) does not indicate any instability. This can easily be seen after the Wick rotation is performed to space with Euclidean signature.  \revisionB{Indeed, $S^{\mu}$ is an axial vector whose components are given by the first line of Eq. (\ref{STT}).} The components of $T_{\mu\nu\rho}$ with one of the indices equal to zero should be multiplied by the imaginary unit during the Wick rotation. Therefore, the spacial components of $S^{\mu}$ are multiplied by the imaginary unit too, while the component $S^0$ remains the same. This rule is inverse compared to the roation of vector torsion $T^{\mu}$. Thus, the signs in front of $T^2$ and $S^2$ in the Euclidean version of Eq. (\ref{S}) are the same and \revision{in Euclidean space} these terms are to be interpreted as the mass terms for the corresponding vector fields \cite{Shapiro}.

Then the  integration over torsion degrees of freedom can be performed for the system that consists of the fermion
coupled to axial torsion $S$ and vector torsion $T$. \revision{For the case of the flavor - independent coupling constants such an integration has been performed in \cite{Freidel:2005sn,Randono:2005up,Mercuri:2006um,Alexandrov,Xue,Alexander1,Diakonov2}. In our case the calculation is similar.} The result of this integration is $S_{eff} = S_{\nabla} + S_E + S_4$, where
\begin{eqnarray}
S_{4}& = & \frac{3 }{32 M_T^2} \int E
\Bigl( \Bigl(-\bar{\Psi}_L \bar{\sigma}_d \Psi_L + \bar{\Psi}_R \sigma_{d} \Psi_R\Bigr)^2\nonumber\\&& -
\Bigl(\bar{\Psi}_L \bar{\sigma}_d B_L \Psi_L + \bar{\Psi}_R B_R  \sigma_{d} \Psi_R\Bigr)^2
\Bigr) d^4x \label{F42}
\end{eqnarray}
while $S_E$ is the effective action that depends on the metric field only.

After the Fierz transformation we arrive at
\begin{eqnarray}
S_{4}& = & \frac{3 }{16 M_T^2} \int E
\Bigl( 2(\bar{\Psi}^{\bf a}_L \Psi_{{\bf b},R})(\bar{\Psi}^{\bf c}_R \Psi_{{\bf d},L}) (\delta_{\bf a}^{\bf d} \delta^{\bf b}_{\bf c} + B^{\bf d}_{{\bf a},L} B^{\bf b}_{{\bf c},R}) \nonumber\\ &&
- (\bar{\Psi}^{\bf a}_L \bar{\Psi}^{\bf c}_L)(\Psi_{{\bf d},L}\Psi_{{\bf b},L})(\delta_{\bf a}^{\bf d} \delta^{\bf b}_{\bf c} - B^{\bf d}_{{\bf a},L} B^{\bf b}_{{\bf c},L}) \nonumber\\&& -
(\bar{\Psi}^{\bf a}_R \bar{\Psi}^{\bf c}_R)(\Psi_{{\bf d},R}\Psi_{{\bf b},R})(\delta_{\bf a}^{\bf d} \delta^{\bf b}_{\bf c} - B^{\bf d}_{{\bf a},R} B^{\bf b}_{{\bf c},R})
\Bigr) d^4x\label{F422}
\end{eqnarray}
Here the Weyl spinor indices $A,B,...$ are omitted \revisionB{and we denote $\bar{\Psi}^A \bar{\Psi}^B \epsilon_{AB} \equiv \bar{\Psi} \bar{\Psi}$, ${\Psi}_A {\Psi}_B \epsilon^{AB} \equiv {\Psi} {\Psi}$.}

\subsection{Auxiliary scalar field}

In principle, all observed masses of quarks and leptons may acquire contributions due to the interaction term Eq. (\ref{F422}). We consider here the simplification, when only the Majorana masses for the three right - handed neutrinos are generated in this way. We assume, that these masses are of the order of the torsion fluctuation scale $M_T$. We introduce the auxiliary scalar field $H$ and arrive at the following interaction term of the effective action:
\begin{eqnarray}
S_{4}  &=&   \int E \Bigl[-\frac{1}{2}\Bigl(H^+[{\nu}_{ R}{\nu}_{R}] + (h.c.)\Bigr) -
\frac{ M_T^2}{g_{\nu}} \, H^+    H \Bigr]
d^4 x,
\label{eff2}
\end{eqnarray}
where we denoted $g_{\nu} = \frac{3}{4}(1-B_{\nu_R}^2)$ while $B_{\nu_R}$ is the diagonal element of matrix $B_R$ that corresponds to the  right - handed neutrino. (We assume for simplicity, that the diagonal elements of $B_R$ that correspond to the right - handed neutrinos are all equal to each other.) {The auxiliary field $H^{{\bf a} {\bf b}}$ carries flavor indices of neutrino ${\bf a}, {\bf b} = e,\mu,\tau$. Sum over these indices is assumed: $H^+ \nu_R \nu_R = \bar{H}_{\bf ab}\nu_R^{\bf a}\nu_R^{\bf b}$ and $H^+ H = \bar{H}_{\bf ab} H^{\bf ab}$.
Vacuum value of $H$ is: $H^{\bf a b} = \delta^{\bf a b} M_{\nu_R}$.  } The coupling constants for the left - handed and the right - handed neutrinos may be different, so that we may choose the values of $B_{\nu_L}, B_{\nu_R}$  in such a way, that the right - handed neutrino is massive while the left handed neutrino is not.

Here and below we restrict ourselves with the case of flat metric and arrive at the effective action for the right - handed neutrino:
\begin{eqnarray}
S_{\nu_R}  &=&   \frac{1}{2}\int  \Bigl(\bar{\nu}_{R} i \nabla_{\mu} \sigma^{\mu}{\nu}_{R} + (h.c.)\Bigr)
d^4 x \nonumber\\&& +\int  \Bigl[-\frac{1}{2}\Bigl(H^+[{\nu}_{ R}{\nu}_{R}] + (h.c.)\Bigr) -
\frac{M_T^2}{g_{\nu}} \, H^+    H \Bigr]
d^4 x
\label{eff2nu}
\end{eqnarray}
We introduce the Nambu - Gorkov spinors
\begin{equation}
N = \left(\begin{array}{c} -i\sigma^2 \bar{\nu}^T_R\\ \nu_R \end{array}\right)
\end{equation}
Those spinors satisfy reality conditions $\bar{N}^T = -i \gamma^2 N$.
Then $\bar{N}\gamma^0N = \Bigl(\nu_R \nu_R + (h.c.) \Bigr)$, $\bar{N}\gamma^0 \gamma^5 N = \Bigl(\nu_R \nu_R - (h.c.) \Bigr)$, and $\bar{N}\gamma^0 \gamma^{\mu} i \nabla_{\mu} N =   \Bigl(\bar{\nu}_{R} i \nabla_{\mu} \sigma^{\mu}{\nu}_{R} + (h.c.)\Bigr)$.  We rewrite the effective action as follows:
\begin{eqnarray}
S_{\nu_R}  &=&   \frac{1}{2}\int  {N}^T \, (-i\, \gamma^2) \gamma^0  \Bigl(i \nabla_{\mu} \gamma^{\mu} - \frac{H^+ + H}{2} -\frac{H^+ - H}{2}\gamma^5\Bigr)N
d^4 x \nonumber\\&& -\int
\frac{ M_T^2}{g_{\nu}} \, H^+    H
d^4 x\label{SeffN}
\end{eqnarray}

The effective action may be expressed through the functional determinant as follows
\begin{eqnarray}
S[H] &=& - i \, {\rm log}\, \int DN \, e^{i S_{\nu_R}[N,h,\omega]}\label{effSh} \\&=& - i\, {\rm log}\,{\rm Pf}\Bigl[(-\gamma^2 \gamma^0)  \Bigl(i \nabla_{\mu} \gamma^{\mu} - {\rm Re}\,  H - i \, {\rm Im}\, H \gamma^5 \Bigr)\Bigr]-\frac{M_T^2}{g_{\nu}}\int
 \,H^+ H
d^4 x\nonumber \\&=& - \frac{i}{2} \, {\rm log} \, {\rm Det}\, \Bigl(i \nabla_{\mu} \gamma^{\mu} - {\rm Re}\, H - i\, {\rm Im} \, H \gamma^5 \Bigr) -\frac{M_T^2}{g_{\nu}}\int
 \,H^+ H
d^4 x\nonumber\\
\end{eqnarray}
(Notice, that the functional determinant is defined up to the constant factor, so that we omit ${\rm Det}^{1/2}\, (-\gamma^2\gamma^0)$.)
Here we used the following expression for the fermionic path integral over the Majorana fermions (i.e. the fermions that satisfy reality conditions):
\begin{equation}
\int D \chi {\rm exp} (- \frac{1}{2}\chi^T {\cal A} \chi) = {\rm Pf} ({\cal A})  = {\rm Det}^{1/2}\, {\cal A},
\end{equation}
where ${\cal A}$ is the skew - symmetric operator (i.e. ${\cal A}^T = - {\cal A} \Leftrightarrow \theta_1^T {\cal A}\theta_2 = - \theta^T_2 {\cal A} \theta_1 $ for bosonic variables $\theta_1,\theta_2$). By ${\rm Pf}(A)$ we denote the Pfaffian. One can easily check, that the operator ${\cal A} = -\gamma^2 \gamma^0  \Bigl(i \nabla_{\mu} \gamma^{\mu} - {\rm Re}\,  H - i \, {\rm Im}\, H \gamma^5 \Bigr)$ is indeed skew symmetric.

\subsection{Regularizations of the NJL model. Low energy effective lagrangian.}
\label{sectcutoff}

\revision{The four - fermion NJL (Nambu - Jona - Lasinio) model defined by Eq. (\ref{eff2}) is not renormalizable. Different regularizations of this model give different results. In the conventional cutoff regularization we are to implement the finite ultraviolet cutoff $\Lambda$ \cite{Miransky:1994vk}. In zeta regularization \cite{zeta,zeta1,zeta2} instead of the cutoff the dimensional parameter $\mu$ appears. This parameter marks the scale of the hidden interaction that results in the effective four - fermion lagrangian. Therefore, the meaning of this parameter is the same as the meaning of the cutoff in the conventional cutoff regularization.  Notice, that the two mentioned regularizations give in general case different effective low energy theories. For the case of the conventional regularization we are to add the set of counter - terms that are to cancel the quadratic divergences. The existence of these counter - terms is necessary to use the $1/N_{\nu}$ approximation. Without such terms the next to leading order approximation gives contributions to various physical quantities that are larger than the one - loop results, and the latter do not have sense (see \cite{VZ2012} and references therein). Assuming, these counter - terms are added and the value of $\Lambda$ is much larger than the value of the induced fermion mass the conventional regularization gives the relations between physical observables (Majoron decay constant, composite scalar boson masses, neutrino Majorana mass etc) the same as given by zeta regularization if we indentify $\mu$ with $\Lambda$. At the same time, the one - loop gap equations (that relate observable neutrino Majorana mass with (non - observable) bare four - fermion coupling constant) are different. This results in the finite renormalization of the bare four - fermion coupling constant when one passes from one discretization to another. The important feature of this finite renormalization is that the bare four - fermion interaction in zeta regularization is repulsive while the bare four - fermion interaction in the conventional regularization is attractive. The obvious advance of zeta regularization is that it does not contain dangerous ultraviolet divergences ab initio, and, therefore, does not require the  introduction of additional counter - terms. The effective theory written in zeta regularization admits working $1/N_{\nu}$ approximation that allows to treat it perturbatively.  That's why we feel this instructive to represent perturbative analysis of the effective theory with action Eq. (\ref{effSh}) using zeta regularization (see Appendix).

The value of $M_T$ governs the fluctuations of torsion itself, the value of $\mu$ is given by the inverse correlation length of the torsion field. The appearance of the value of $\mu$ different from the value of $M_T$ may be understood when the simple toy model for the dynamical torsion is considered. For simplicity let us consider vector torsion $T^{\mu}$ with the following lagrangian that contains the kinetic term:
\begin{equation}
L =  - \frac{2 Z_T}{6} \partial_{[\nu}T_{\mu]}\partial^{[\nu}T^{\mu]} + \frac{2M_T^2}{3}T^{\mu}T_{\mu}\label{ST-0}
\end{equation}
Here $Z_T$ is the wave function renormalization constant for vector torsion.  The generated correlation length for the field  $T$  is given by: $\zeta_T = \frac{Z_T^{1/2}}{M_T}$.
Then the natural cutoff is given by
\begin{equation}
\mu \approx  \frac{M_T}{Z_T^{1/2}} \label{M_TL}
\end{equation}
The one - loop results presented in Appendix prompt, that the value of $\mu$ may be much larger, than $M_T$.
The simple explantation of this is that the dimensionless constant $Z_T$ is generated dynamically, and may be much less than unity.
In general case various terms that depend on the derivatives of axial and vector torsion as well as the tensor component $q_{abc}$ of Eq. (\ref{STT})  were considered in Ref. \cite{Diakonov2}. The situation in general case is more complicated, the issues of the stability of the corresponding quantum system are to be addressed. However, this is out of the scope of the present paper. It is only important for us, that similar to the toy model with action of Eq. (\ref{ST-0}) the torsion correlation length $\sim \frac{1}{\mu}$ may be much smaller than the value of $1/M_T$ due to the small values of the effective constants standing at the kinetic terms.}

The low energy effective lagrangian for the interaction of the right - handed neutrinos with  scalar fields  $ H^{\bf a b} ({\bf a}, {\bf b} = e,\mu, \tau)$ can be written as follows:
\begin{eqnarray}
L_{eff} &=&  \frac{F^2_{\nu}}{4 N_{\nu} M_{\nu}^2} \nabla \bar{H}^{\bf ab} \nabla H_{\bf ab} - V(H) -   \frac{1}{2}\Bigl(\nu_{\bf a}\nu_{\bf b} \bar{H}^{\bf ab} + (h.c.) \Bigr)\label{Seffsimple}
\end{eqnarray}

The potential for the scalar fields $V$ has its minimum at $H^{\bf ab} = M_{\nu}\delta^{\bf ab}$. In the analysis of Eq. (\ref{Seffsimple}) we take into account that the metric field is frozen and flat. This means that $E=1$, and that the pure fermion action is to be taken in its simplest form of Eq. (\ref{Ssimple}). Bosonic part of the effective lagrangian of the form of Eq. (\ref{Seffsimple}) appears, in particular,  in the leading order of $1/N_{\nu}$ approximation, i.e. the one loop approximation (see Eq. (\ref{SH}) of Appendix).

\subsection{Seesaw and light neutrino masses}

As it was mentioned above, we assume, that all neutrinos acquire Dirac masses of the order of the electron mass. Therefore, we have the classical type I seesaw \cite{Hernandez:2010mi}. In the basis ${\cal N} = (\nu^c_L, \nu_R)^T$ (where $\nu_L^c = i\sigma^2 \bar{\nu}_L$) there is the mass matrix
\begin{equation}
{\bf M}_{\nu} = \left(\begin{array}{cc} 0 & m_{\nu}\\
m_{\nu} &   M_{\nu_R} \end{array}\right)\label{MASSSEESAW}
\end{equation}
The overall mass term is $\frac{1}{2}{\cal N} {\bf M}_{\nu} {\cal N} + (h.c.)$, where $(h.c.)$ means hermitian conjugation that implies $\nu_{R,L} \rightarrow \bar{\nu}_{R,L}$ and vice versa. For simplicity we assume, that the three Dirac masses $m_{\nu}$ are equal to each other and the three Majorana masses $M_{\nu_R}$ are also equal.
 The diagonalization gives the three heavy neutrinos with Majorana masses $\sim M_{\nu_R}$ and three light neutrinos with Majorana masses
 \begin{equation}
 M^{light}_{\nu} \approx m_{\nu} \frac{m_{\nu}}{ M_{\nu_R}}
\end{equation}

Our assumptions that the mass of the right - handed neutrinos $M_{\nu_R}$ is not smaller, than $1$ TeV, and that the Dirac neutrino mass $m_{\nu}$ is of the order of the electron mass $m_e$ allow us to estimate
$ M^{light}_{\nu}  \le  0.25$ \, eV.

\subsection{Majorons and their possible impact on the experimental results}

Majorons (Goldstone bosons that appear during the spontaneous breaking of lepton number) are massless in the given model. However, their coupling to ordinary matter is such small, that they may escape experimental detecting \cite{Schechter:1981cv, Diaz:1998zg}. The explanation of this is that the right - handed neutrinos (like the ones that acquire large Majorana masses due to the mechanism suggested here) do not interact with the other particles of the Standard Model. The resulting states that diagonalize the mass matrix Eq. (\ref{MASSSEESAW}) are composed of $\nu_L^c, \nu_L$, and  $\nu_R, \nu^c_R$ as
\begin{equation}
\nu^{\rm light}_L \approx \nu_L + \epsilon \nu^c_R, \quad \nu^{\rm heavy}_R \approx \nu_R + \epsilon \nu^c_L,
\end{equation}
where the small parameter is
\begin{equation}
\epsilon = m_{\nu} \Bigl(M_{\nu_R}\Bigr)^{-1}\sim 10^{-6}
\end{equation}
Majoron is composed of $\nu_R$. Therefore, its coupling to the observed light neutrino is due to the transition between $\nu^{\rm light}_L$ and $\nu_R$. Thus this coupling is suppressed by $\epsilon^2\sim 10^{-12}$. For the limits on the coupling between Majorons and the SM fermions see \cite{MajoronCouplings}.
Such a small coupling between usual matter and majorons allows to consider the latter as a candidate for the dark matter (see, for example, \cite{MajoronDarkMatter} and references therein).

The supposed value for the Majorana mass of heavy neutrino of the order of $M_T$ (that may be not far from the Higgs boson mass) prompts that the Higgs boson mass and the mass of heavy neutrino may have the common origin. This does not mean, that the origin of the Higgs boson mass is necessarily related to the interaction with torsion. This means rather that the interaction between torsion and neutrino that has led to the appearance of Majorana mass is related somehow to the unknown interactions that presumably generate $M_H \approx 125$ GeV. In this speculation we imply that the recently discovered $125$ GeV Higgs boson is composite. See, for example, the recent paper \cite{Z2014PRD}, where it is suggested that due to the new strong interaction the observed $125$ GeV Higgs boson may appear to be composed of all known SM fermions. One may speculate, that such a hidden strong interaction between the SM fermions is related somehow to the interaction between the SM fermions and torsion. If so, it is natural to suppose, that there is a coupling between Majorons $J$ and the Higgs boson $h$ due to their assumed common origin (or, the origins that are related). This coupling is not necessarily small unlike the direct coupling between Majorons and the other particles of the Standard Model. In particular, the processes like $h \rightarrow JJ$ may generate the invisible decay of the Higgs boson. The present experimental constraints on the corresponding branching ratio \cite{HIGGSINVISIBLE} are rather light: the upper bound on the invisible decay branching ratio is about $75 \%$. Therefore, presently, the experimental constraint on the coupling between Majorons and the Higgs boson is almost absent.

\section{Conclusions and discussions}
\label{concl}

In this paper we suggest that the dynamics behind the formation of the Majorana masses for the right - handed neutrinos is related to the coupling of the fermions to quantum gravity with torsion. We assume, that there are two major scales in quantum gravity. The conventional scale of the order of Plank mass corresponds to the fluctuations of metric while the other scale $M_T$ corresponds to the fluctuations of torsion. (Such a situation was already discussed previously, see Ref. \cite{Shapiro} and references therein.) The latter scale may be as low as $1$ TeV that does not contradict to the present experimental constraints Ref. \cite{Shapiro}. Therefore, we are left with the gauge theory of the Lorentz group. It has been demonstrated in the present paper, that the attractive forces are formed not only between fermion and antifermion of different chiralities, but also between the fermions of the same chiralities but with different directions of spin. This may result in the formation of the Majorana masses.

The key ingredient of our model is the nontrivial matrices of coupling constants $B_L$ and $B_R$ entering  Eq. (\ref{Sf}) and the corresponding action for the left - handed spinors. These matrices give the different strength of the non - minimal coupling to torsion of different fermions. In particular, left and right - handed fermions may interact with torsion with different strength. In principle, via an appropriate choice of coupling constants the essential contributions to all masses of quarks and leptons may be reproduced. However, we concentrate on the appearance of the Majorana masses for the right - handed neutrinos only.

The main output of the present paper is the explanation of the appearance of small neutrino masses  $\le 0.25$ eV. Those masses appear due to the type I seesaw mechanism, in which the right - handed neutrinos acquire the masses of the order of $M_T$. The secondary output follows from the existence of massless Majorons (Goldstone bosons)\footnote{It is worth mentioning, that in addition to massless Majorons the model predicts the appearance of scalar bosons composed of the right - handed neutrinos. The corresponding mass is calculated in one - loop approximation and is given by $2M_{\nu_R}$, where $M_{\nu_R}$ is the mass of the heavy neutrino.}. The Majorons interact with SM fermions only weakly, so that they may escape the direct observations. As a result, they may play the role of Dark Matter \cite{MajoronDarkMatter}. Moreover, assuming that the recently discovered $125$ GeV Higgs boson is composite \footnote{Say, the hidden strong interaction between the SM fermions may lead to the compositeness of the Higgs boson as explained in \cite{Z2014PRD}. This approach generalizes the original top - quark condensation models \cite{Miransky} but does not inherit their main difficulties.} and that the corresponding hidden dynamics correlates somehow with the interaction between the SM fermions and torsion, we come to the conclusion, that the processes $h \rightarrow JJ$ are not suppressed and may result in the invisible decay of the Higgs boson. Notice, that the present experimental constraint on the invisible branching ratio of the Higgs boson decay is $75\%$ at $95\%$ CL (see, for example, the results by ATLAS \cite{HIGGSINVISIBLE}).

\section*{Acknowledgements}

This paper has appeared as a result of the discussions with V.A.Miransky. The author is grateful to him for these discussions, for careful reading of the manuscript, remarks and corrections. The author benefited from discussions with D.I. Diakonov and from useful correspondence with I.L.Shapiro. The continuous support, sharing ideas, and discussions with G.E.Volovik are kindly acknowledged. The work  is  supported by the Natural Sciences and Engineering Research Council of
Canada.

\section*{Appendix. NJL model in zeta regularization.}

\subsection{Evaluation of the fermion determinant}

The simple way to construct the effective theory in such a way, that the dangerous quadratic divergences are cancelled is to use zeta regularization \cite{zeta,McKeon}. In this regularization the ultraviolet divergences are absent at all. As a result, the $1/N$ expansion may always be applied to the theory. However, the price for this is the finite renormalization of the coupling constants. In particular, the interaction that is repulsive at the level of bare lagrangian becomes attractive after renormalization and causes the condensation of fermions and the appearance of masses.

Here we use the method for the calculation of various Green functions using zeta - regularization developed in \cite{McKeon,McKeon1,McKeon2}.
In order to calculate the fermion determinant we perform the rotation to Euclidean space - time. It corresponds to the change: $t \rightarrow -i x^4, \bar{\psi} \rightarrow i \bar{\psi}, \gamma^0 \rightarrow \Gamma^4, \gamma^k \rightarrow i \Gamma^k (k = 1,2,3)$. (The new gamma - matrices are Euclidean ones.) We start from Eq. (\ref{effSh}) and set ${\rm Im}\, H = 0$. Moreover, we consider only the trace modes of $H$ so that it is implied to be proportional to unity matrix $H^{\bf ab} = (M_{\nu}+h) \, \delta^{\bf ab}$.
The resulting Euclidean  functional determinant has the form:
\begin{eqnarray}
Z[h] &=& {\rm Det}^{1/2} \Bigl[\hat{P}\Gamma + iM_{\nu_R} + i h  \Bigr] \nonumber\\ &=& \int DN e^{- \frac{1}{2}\int \bar{N}\Gamma^4\Bigl[\hat{P}\Gamma + iM_{\nu_R} + i h  \Bigr]N d^4x}
\end{eqnarray}
Here $M_{\nu_R}$ is the neutrino Majorana mass.
 The transformation $N \rightarrow \Gamma^5 N , \bar{N}\Gamma^4 \rightarrow -\bar{N}\Gamma^4\Gamma^5$ results in $Z[h] = {\rm Det}^{1/2} \Bigl[\hat{P}\Gamma + i(M_{\nu_R} + h)  \Bigr] = {\rm Det}^{1/2} \Bigl[\hat{P}\Gamma - i(M_{\nu_R} + h)  \Bigr]$. Therefore,
\begin{equation}
Z[h] = \Bigl({\rm Det} \Bigl[\hat{P}\Gamma + i(M_{\nu_R} + h)  \Bigr] \Bigl[\hat{P}\Gamma - i(M_{\nu_R} + h)  \Bigr]\Bigr)^{1/4}
\end{equation}
We
get the fermionic part of the Euclidean effective action:
\begin{equation}
S_f[h] = - \frac{1}{4}{\rm Tr} \, {\rm log} \Bigl[\hat{P}^2 + M_{\nu_R}^2 + (2 M_{\nu_R} h +h^2 -  \Gamma [\partial,h]) \Bigr]
\end{equation}
We denote $A = \hat{P}^2 + M_{\nu_R}^2$, and $V = 2 M_{\nu_R} h +h^2 -  \Gamma [\partial, h]$.
In zeta - regularization we have:
\begin{equation}
S_f[h] = \frac{1}{4} \partial_s \frac{1}{\Gamma(s)} \mu^{2s} \int dt t^{s-1} {\rm Tr} \,{\rm exp}\Bigl[ - (A + V) t\Bigr]
\end{equation}
At the end of the calculation $s$ is to be set to zero.  Here the dimensional parameter $\mu$ appears. We identify this dimensional parameter with the working scale of the interaction that is responsible for the formation of the Majorana mass. Further we expand:
\begin{equation}
{\rm Tr} \,{\rm exp}\Bigl[ - (A + V) t\Bigr] = {\rm Tr} \,\Bigl(e^{ - A t} + (-t) e^{ - A t} V + \frac{(-t)^2}{2}\int_0^1 du  e^{ - (1-u)A t}V e^{ - u A t} V + ...\Bigr)\label{expansion}
\end{equation}
The further derivation follows closely the one given in Appendix of \cite{Z2014PRD}. The final answer is to be analytically continuated to space - time of Minkowski signature.

\subsection{Effective action up to the terms quadratic in $h$.}
The total one - loop  effective action (i.e. the action up to the terms quadratic in $h$) receives the form:
\begin{equation}
S[h] =  \int {d^4x} \Bigl[- \frac{M_T^2N_{\nu}}{g_{\nu}} (M_{\nu_R}+h)^2 -  \frac{1}{2}\hat{C} M_{\nu_R}^2 (M_{\nu_R}+h)^2 +  \frac{1}{2}h(x)\, Z_h^2\, (\hat{p}^2 - M_H^2) w(\hat{p}^2) h(x) \Bigr],\label{Sh}
\end{equation}
where
\begin{eqnarray}
&&w(p^2) = \frac{1}{{\rm log}\, \frac{\mu^2}{M_{\nu_R}^2}}\int_0^1 du \,{\rm log}\, \frac{\mu^2}{-p^2 u(1-u) + M_{\nu_R}^2}\nonumber\\
 && Z_h^2  =   \frac{N_{\nu}}{16 \pi^2}{\rm log}\, \frac{\mu^2}{m_t^2}, \quad  M_H^2 = 4  M_{\nu_R}^2, \quad  \hat{C} = \frac{N_{\nu}}{8 \pi^2 } {\rm log}\frac{\mu^2}{M_{\nu_R}^2} \label{zw}
\end{eqnarray}
Here the imaginary part of the effective action that corresponds to the decay of the composite scalar (Higgs) boson into two right - handed neutrinos appears for $p^2 \ge 4 M_{\nu_R}^2$. Notice, that for $|p^2| \ll \mu^2$ we have $w(p^2) \approx 1$.
The value $M_{\nu_R}$ of the neutrino mass satisfies gap equation $\frac{\delta}{\delta h} S[h] = 0$.
 Then we relate bare parameter $B_{\nu}$ to the scale $\mu$ and the generated neutrino mass $M_{\nu_R}$:
\begin{equation}
M_T^2  = -\frac{3}{4}(1-B_{\nu_R}^2)\frac{1}{16 \pi^2 } {M_{\nu_R}^2}\, {\rm log}\frac{\mu^2}{M_{\nu_R}^2}
\end{equation}
The minus sign here means, that the bare four - fermion interaction of Eq. (\ref{eff2nu}) at the high energy scale $\mu \gg M_{\nu_R}$  is repulsive.  However, the appearance of the vacuum average for the auxiliary field $R_{\nu}$ means, that this repulsive interaction is subject to finite renormalization: at low energies $\ll \mu$, where the neutrino mass is formed, the renormalized interaction between the neutrinos is attractive. In zeta regularization bare parameter $|B^{\rm zeta}_{\nu}| $ should be larger than unity. This distinguishes essentially zeta regularization from the conventional cutoff regularization, in which bare four - fermion interaction is attractive and bare parameter $|B^{\rm conventional}_{\nu}|$ should be less, than unity. This means, that bare parameters of the two regularizations do not coincide and are related by finite renormaization. This finite renormalization is given by
$
([B^{\rm zeta}_{\nu}]^2 -1 ) \Bigl(-0+\frac{M_{\nu_R}^2}{\mu_{}^2}\, {\rm log} \, \frac{\mu_{}^2}{M_{\nu_R}^2}\Bigr)\approx (1-[B^{\rm conventional}_{\nu}]^2)\Bigl(1-\frac{M_{\nu_R}^2}{\Lambda_{}^2}\, {\rm log} \, \frac{\Lambda_{}^2}{M_{\nu_R}^2}\Bigr)
$.
 (Here we denote the Ultraviolet cutoff of the conventional regularization by $\Lambda$ and the scale parameter of zeta regularization by $\mu$.)

 One can see, that there is only one pole of the propagator for the field $h$. It gives the mass for the composite scalar boson (trace Higgs mode):
\begin{eqnarray}
&& M_H \approx 2 M_{\nu_R}
\end{eqnarray}

\subsection{Reconstruction of the whole effective action. Majoron decay constant. }
\label{sectmu}

We may consider Eq. (\ref{Sh}) as the low energy effective action at most quadratic in the scalar field. There exists the phenomenological effective action written in terms of the field $H$ with lagrangian of the form of Eq. (\ref{Seffsimple}). It should contain the potential $V(H)$ that provides the nonzero vacuum average $\langle H^{\bf ab} \rangle = M_{\nu_R} \delta^{\bf ab}$. Comparing the coefficient at the kinetic term of this effective action with that of Eq. (\ref{Sh}) we arrive at the following expression for its bosonic part
\begin{eqnarray}
S[H] =  \int {d^4x}  \Bigl[  \frac{Z_h^2 }{2N_{\nu}} \, \nabla\bar{H}^{\bf ab}(x)\, \nabla H_{\bf ab}(x) - V(H)\Bigr]\label{SH}
\end{eqnarray}
(Recall, that for $|p^2|\ll \mu^2$ we have $w(p^2) \approx 1$.) From here we derive the Majoron decay constant $F_{\nu}$ (that is the analogue of the technipion decay constant) as the coefficient at the term $\frac{\nabla \bar{H}^{\bf ab}\nabla {H}_{\bf ab}}{4\langle \, \bar{H}^{\bf ab}  \,H_{\bf ab}\rangle}$:
\begin{equation}
{F_{\nu}^2} \approx  2 Z_h^2 M_{\nu_R}^2 = \frac{ N_{\nu}}{8 \pi^2} M_{\nu_R}^2 {\rm log}\, \frac{\mu^2}{M_{\nu_R}^2}\label{eta_}
\end{equation}

\end{document}